\documentstyle[aps,epsf,preprint]{revtex}
\begin{document}
\pagestyle{empty}
\begin{flushright}
FERMILAB-PUB-99/265-E\\
\today
\end{flushright}
\footnotetext[1]{Submitted to Phys. Rev. Lett.}

\begin{center}
{\Large\bf A New Limit on CPT Violation
\footnotemark[1]}
\end{center}

\begin{center}
\renewcommand{\baselinestretch}{1}
S. Geer, J. Marriner, M. Martens, R.E. Ray, J. Streets, W. Wester \\
{\it Fermi National Accelerator Laboratory, Batavia, Illinois 60510}\\
\vspace{0.2cm}
M. Hu, G.R. Snow \\
{\it University of Nebraska, Lincoln, Nebraska 68588}\\
\vspace{0.2cm}
T. Armstrong \\
{\it Pennsylvania State University, University Park, Pennsylvania 16802}\\
\vspace{0.2cm}
C. Buchanan, B. Corbin, M. Lindgren, T. Muller$^{\dag}$ \\
{\it University of California at Los Angeles, Los Angeles, California 90024} \\
\vspace{0.2cm}
R. Gustafson \\
{\it University of Michigan, Ann Arbor, Michigan 48109}\\
\end{center}

\vspace{0.1in}
\begin{center}
(The APEX Collaboration)
\end{center}
\begin{abstract}
\nopagebreak
A search for antiproton decay 
has been made at the Fermilab Antiproton Accumulator. Limits are 
placed on fifteen antiproton decay modes. The results are used 
to place limits on the characteristic mass scale $m_X$ 
that could be associated with CPT--violation accompanied by 
baryon number violation.
\end{abstract}


\clearpage
\pagestyle{plain}
\setcounter{page}{1}
%
%

The CPT theorem requires that the proton ($p$) and antiproton
($\overline{p}$) lifetimes are equal.
Searches for $p$ decay have yielded lower limits on
the $p$ lifetime $\tau_p > O(10^{32})$~yr~\cite{protondecay}.
A search for $\overline{p}$ decay with a short lifetime
($\tau_{\,\overline{p}} << \tau_p$) tests the CPT theorem.
In this paper we summarize results from a search for $\overline{p}$ decay
at the Fermilab Antiproton Accumulator, and discuss limits on 
the characteristic mass scale $m_X$ associated with CPT--violation. 
For a mass dimension--5 CPT violating 
operator these limits approach the Planck scale.

CPT invariance is one of the most fundamental symmetries
of modern physics. There have been a variety of
searches~\cite{protondecay,cptlimits,ion,k0limits} for CPT violation based upon
comparing particle and antiparticle masses, lifetimes, and 
magnetic moments. For example, 
the $p$ and $\overline{p}$ masses have been shown to be equal 
with a precision of a few parts in $10^{8}$~\cite{ion}, while 
the particle and antiparticle masses in the
neutral kaon system have been shown to be equal to about 
one part in $10^{18}$~\cite{protondecay,k0limits}. 
A search for $\overline{p}$ decay
complements these CPT tests by providing a search for CPT violation
accompanied by a violation of baryon number. Indeed,
since the $\overline{p}$ is the only very long--lived
antiparticle that could in principle decay into other known particles
without violating charge conservation, a search for $\overline{p}$ decay
provides a unique test of the CPT theorem, and a unique test of the
intrinsic stability of antimatter. 

The sensitivity of a $\overline{p}$ decay
search to the presence of a CPT violating interaction has been
characterized by considering a
dimension-$n$ CPT-violating quantum field operator 
($n > 4$) with characteristic
mass scale $m_X$. Dimensional analysis then provides
the estimate~\cite{dallas} 
$m_p \tau_{\,\overline{p}} \sim [m_p/m_X]^{2n-8}$, yielding :
\begin{equation}
m_X/m_p\sim [4.5\times 10^{38}\cdot
\tau_{\,\overline{p}} /10{\rm\ Myr}]^{1/(2n-8)} \; .
\label{mxlimit}
\end{equation}
For a given lower limit on $\tau_{\,\overline{p}}$ 
the implied lower limit on $m_X$ is most stringent for $n$ = 5.
Note that if $m_X$ is at the Planck scale $(1.2\times 10^{19}$ GeV/$c^2)$ and 
$n$ = 5, the expected $\tau_{\,\overline{p}}$ 
would be $\sim$ 10 Myr.

The most stringent lower limit on $\tau_{\,\overline{p}}$
has been obtained~\cite{crprl} from a comparison of recent measurements
of the cosmic ray $\overline{p}$ flux~\cite{crays}
with predictions based on expectations
for secondary production of antiprotons in the interstellar medium.
The agreement between the observed and predicted rates implies
that $\tau_{\,\overline{p}}$ is not small compared to T/$\gamma$, where
T is the $\overline{p}$ confinement time within the galaxy ($\sim 10^7$~yr)
and $\gamma$ is the Lorentz factor for the observed antiprotons.
After taking into account the uncertainties on the
relationship between the interstellar
$\overline{p}$ flux and the flux observed at the Earth,
at 90\% C.L. the limit $\tau_{\,\overline{p}} > 8 \times 10^5$~yr
has been reported~\cite{crprl}.
This indirect limit is not valid if current models of $\overline{p}$
production, propagation, and interaction in the interstellar medium are
seriously flawed. For example, 
within the minimal supersymmetric extension to the standard model, 
a significant cosmic ray $\overline{p}$ flux can be produced
by cold dark matter neutralino annihilation. In this scenario, if 
the depletion of the spectrum due to $\overline{p}$ decay is compensated 
by additional contributions from neutralino annihilation, it has been 
claimed~\cite{susy} that the current cosmic ray data can accommodate 
$\overline{p}$ lifetimes as low as $\le 10^5$~years.

Laboratory searches for $\overline{p}$ decay have, to date, provided less
stringent limits on $\tau_{\,\overline{p}}$. However, these limits do not
suffer from large model dependent uncertainties.
The most stringent published laboratory limit on inclusive
$\overline{p}$ decay has been obtained from a
measurement of the containment lifetime of $\sim 1000$ antiprotons
stored in an ion trap, yielding
$\tau_{\,\overline{p}} > 3.4$~months~\cite{ion}.
The sensitivity of laboratory $\overline{p}$ decay searches can be
improved by looking for explicit $\overline{p}$ decay modes
at a $\overline{p}$ storage ring.
Angular momentum conservation requires that a decaying $\overline{p}$ would
produce a fermion (electron, muon, or neutrino) in the final state.
A search for explicit $\overline{p}$
decay modes with an electron in the final state was made
by the T861 experiment at the Fermilab Antiproton Accumulator.
The T861 search yielded the 95\% C.L. limits~\cite{t861}:
$\tau_{\,\overline{p}}/B(\overline{p} \rightarrow e^-\gamma) > 1848$~yr,
$\tau_{\,\overline{p}}/B(\overline{p} \rightarrow e^-\pi^0) > 554$~yr,
$\tau_{\,\overline{p}}/B(\overline{p} \rightarrow e^-\eta) > 171$~yr,
$\tau_{\,\overline{p}}/B(\overline{p} \rightarrow e^-K^0_S) > 29$~yr, and
$\tau_{\,\overline{p}}/B(\overline{p} \rightarrow e^-K^0_L) > 9$~yr.

Following the T861 results, the APEX experiment~\cite{nim}
was designed to enable a more sensitive search for $\overline{p}$ decay. 
The APEX detector, located in a straight section of the 
474~m circumference Fermilab Antiproton Accumulator ring,  
was designed to identify $\overline{p}$ decays within 
a 3.7~m long evacuated decay tank. Particles exiting the tank 
at large angles to the circulating $\overline{p}$ beam 
traversed a 96~cm diameter 1.2~mm thick 
stainless steel vacuum window. The experiment was optimized to detect 
a single energetic charged track (electron or muon), 
originating from the beam, and accompanied by one or more neutral pions or 
photons. 
A detailed description of the detector can be found in Ref.~\cite{nim}. 
In brief, the detector consists of 
(i) An upstream system of scintillation counters arranged 
around the 10~cm diameter beam pipe, located upstream of 
the tank, and used to reject upstream interactions.
(ii) Three planes of horizontal and three 
planes of vertical scintillation counters downstream of the tank. 
Each plane consisted of two $50 \times 100 \times 1.27$~cm$^3$ counters. 
The last counter planes 
were downstream of a 2.3 radiation length lead wall, providing a preradiator 
to aid in identifying electrons and photons. The first and second counter 
planes, upstream of the lead, provided pulse height information used to 
count the traversing charged particles. 
(iii) A lead--scintillator sampling electromagnetic 
calorimeter~\cite{ref:fcal} constructed from 144 rectangular 
$10 \times 10$~cm$^2$ modules that are 14.7 radiation lengths deep, 
arranged in a $13 \times 13$ array with 6 modules 
removed from each of the 
four corners, and the central module removed to allow passage of the 
beam pipe.
(iv) A tail catcher (TC) downstream of the calorimeter 
consisting of a 20~cm deep lead wall followed by two planes of 
scintillation counters.
(v) A muon telescope (MT) downstream of the TC, 10
nuclear interaction lengths deep, 
and aligned to point towards the center of the decay tank. 
The MT consists of a sandwich of five iron plates and five 
$30 \times 30$~cm$^2$ scintillation counters, and was used to 
identify penetrating charged particles (muon candidates).
(vi) A tracking system consisting of three planes of horizontal 
and three planes of vertical 2~mm diameter 
scintillating fibers downstream of the tank and upstream of the preradiator 
lead. These detectors provided three space points 
along the track trajectory with typical residuals of 620~$\mu$m in the 
directions transverse to the beam direction, enabling the 
origin of tracks emerging from the decay tank to be reconstructed 
with an rms precision given by $\sigma_z = 12$~cm. 

The APEX experiment took data 
at times when there were typically $10^{12}$ 
antiprotons circulating in the Accumulator ring with a
central $\overline{p}$ momentum of $8.90 \pm 0.01$~GeV/c
($\gamma = 9.54 \pm 0.01$). 
A measure of the sensitivity of the APEX data sample is given by:
\begin{equation}
S \;\equiv\; \frac{1}{\gamma} \int{N_{\overline{p}}\,(t)\,dt}
\;=\; (3.31 \pm 0.03) \times 10^{9}\;\;yr,
\end{equation}
where N$_{\overline{p}}\,(t)$ is the number of circulating antiprotons at time
$t$, the integral is over the live-time of the
experiment, and the uncertainty arises from the precision with which the
time dependent beam current was recorded.

Energetic particles passing through the detector during Accumulator 
operation predominantly arise from interactions of the
$\overline{p}$ beam with the residual gas in the decay tank or with
material surrounding the beam. To suppress these backgrounds, 
and select candidate $\overline{p} \rightarrow \mu^{-} X$ and 
$\overline{p} \rightarrow e^{-} X$ 
decays, signals from the 
calorimeter and the scintillation counters were used to form an MT 
trigger which recorded 
1.2 $\times 10^{6}$ events, and calorimeter triggers which 
recorded 37.8 million events. 
Offline, events were selected for analysis if they had 
(i) a single charged track reconstructed in the tracking system that pointed 
back to the beamline within the fiducial volume of the decay tank, and 
(ii) a pattern of hits in the scintillation counters and calorimeter cells 
consistent with a 2-body or 3-body $\overline{p}$ decay topology. Events that 
passed these requirements were then required to be kinematically consistent
with the particular decay channel being considered (see Table~1). 
A detailed description of the triggers, data taking, and analysis can be 
found in Refs.~\cite{nim,theses,muonprl,prd}. 
No statistically significant $\overline{p}$ decay signal was observed. 
For a specific decay mode with branching ratio $B$ this null result 
can be used to place a limit on $\tau/B$ which is given in years by:
\begin{equation}
\tau_{\,\overline{p}}/B(\overline{p} \rightarrow \mu^-X) \; > \;
- \; \frac{1}{ln(1-N_{max}/\epsilon S)} \; ,
\end{equation}
where $\epsilon$ 
is the calculated fraction of decays taking place uniformly
around the accumulator ring that pass the trigger and event
selection requirements. 
The upper limit on the number of signal events $N_{max}$ is given by 
the prescription of Ref.~\cite{cousins}:
\begin{equation}
N _{max} \; = \; \mu_{max} \times (1 + \mu_{max}~\sigma_r^2/2) \; ,
\end{equation}
where $\sigma_r \equiv \sigma_\epsilon/\epsilon$, 
$\sigma_{\epsilon}$ is the systematic uncertainty on
$\epsilon$, and 
$\mu_{max}$ is the 90\% C.L. upper limit corresponding to the observation of 
$N$ events that pass the trigger and selection requirements. 
A detailed Monte Carlo simulation of the detector geometry 
and response has been used to calculate $\epsilon$ and $\sigma_r$.

\begin{figure}
\epsfxsize6.in
\centerline{\epsffile{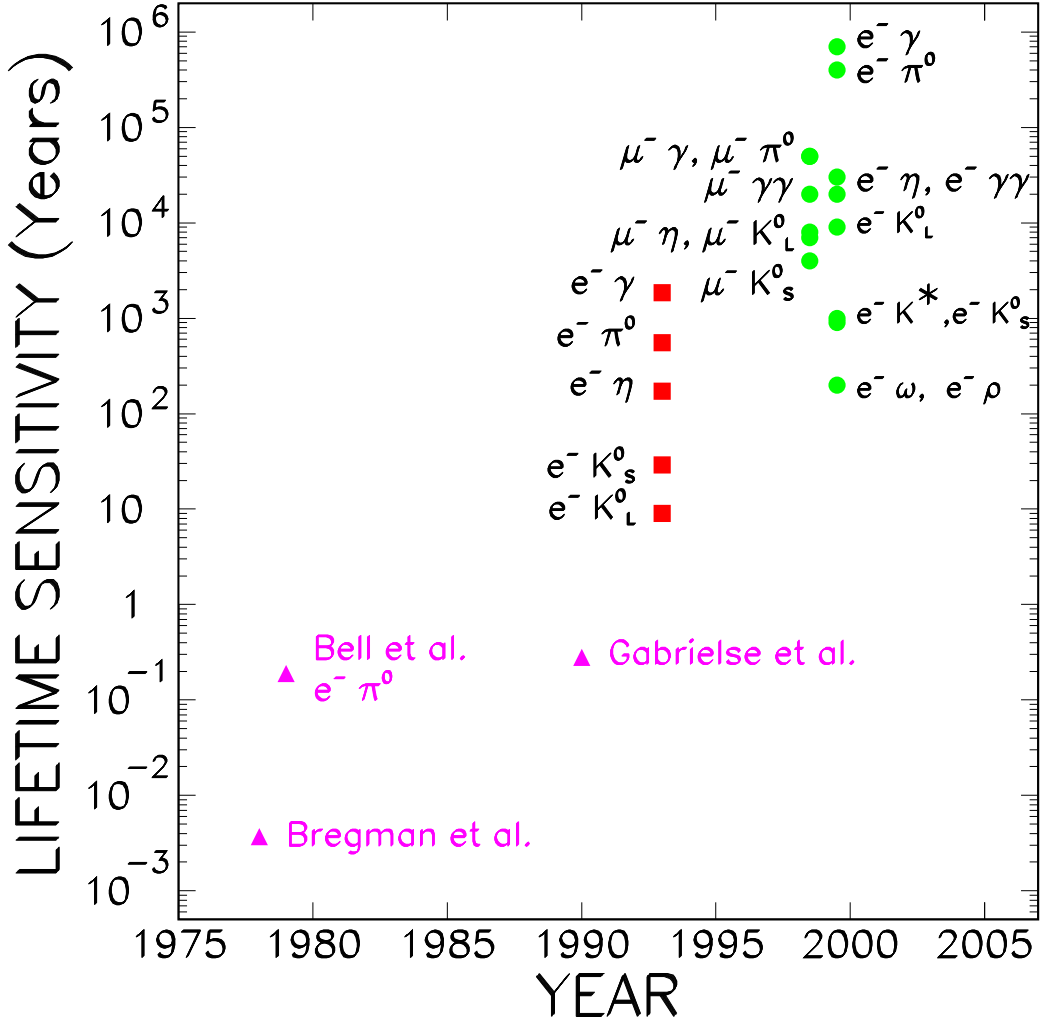}}
\vspace{0.5cm}
\caption{Upper limits on the antiproton lifetime. APEX limits (circles), 
T861 limits (boxes), and limits \protect\cite{ion,bell,bregman}
prior to the T861 experiment (triangles) 
are shown as a function of the publication year.}
\label{limits_fig}
\end{figure}

Table~1 lists the 
resulting limits on $\tau/B$ for 15 decay modes 
in which there is an electron or muon in the final state. 
Our results are shown in Fig.~\ref{limits_fig} to be 
significantly more stringent than previous limits~\cite{cptlimits}. 
In particular, we place the first explicit 
limits on the muonic decay modes of the $\overline{p}$, and the first 
limits on the decay modes $e^-\gamma\gamma$, $e^-\rho$, $e^-\omega$, 
and $e^-K^{0^\star}$. 
Our most stringent limit is on the decay 
$\overline{p} \rightarrow e^-\gamma$ for which we find 
$\tau/B > 7 \times 10^5$~yr (90\% C.L.). 
Table~2 summarizes the limits on the CPT--violating mass scale derived 
from Eq.~(1), listed as a function of the mass dimension--$n$ 
for $5 \le n \le 9$. 
Note that for the most sensitive decay modes the lower limit on 
$m_X$ is O($10^{18}$)~GeV/c$^2$ for $n = 5$. For $n = 6$ the limits 
are in the range $10^8 - 10^9$~GeV/c$^2$ for all 15 decay modes. 
Even for $n = 8$ the limits on $m_X$ are still at the 10~TeV/c$^2$ scale. 
These are unique limits on the presence of CPT--violation 
accompanied by a violation of baryon number.

The APEX experiment was performed at the Fermi National Accelerator 
Laboratory, which is operated by Universities Research Association,
under contract DE-AC02-76CH03000 with the U.S. Department of
Energy. We are grateful to Dallas Kennedy for comments on the paper.
%
%

\clearpage

\begin{table}
\centering
\caption{Summary of results: 90\% C.L. limits on $\tau / B$ for 15 antiproton 
decay modes.}
\vspace{0.2cm}
\begin{tabular}{cccc}
Decay Mode &$\tau/B$ Limit & Decay Mode &$\tau/B$ Limit\\
           & (years) &  & (years) \\
\hline
$\mu^-\gamma$       & $5 \times 10^4$ & $e^-\gamma$      & $7 \times 10^5$\\
$\mu^-\pi^0$        & $5 \times 10^4$ & $e^-\pi^0$       & $4 \times 10^5$\\
$\mu^-\eta$         & $8 \times 10^3$ & $e^-\eta$        & $2 \times 10^4$\\
$\mu^-\gamma\gamma$ & $2 \times 10^4$ & $e^-\gamma\gamma$ & $2 \times 10^4$\\
$\mu^-K^0_L$        & $7 \times 10^3$ & $e^-K^0_L$       & $9 \times 10^3$\\
$\mu^-K^0_S$        & $4 \times 10^3$ & $e^-K^0_S$       & $9 \times 10^2$\\
                    &                 & $e^-\rho$        & $2 \times 10^2$\\
                    &                 & $e^-\omega$      & $2 \times 10^2$\\
                    &                 & $e^-K^{0^\star}$ & $1 \times 10^3$\\
\end{tabular}
\protect\label{results_tab}
\end{table}

\begin{table}
\centering
\caption{Summary of limits on the CPT--violating scale $m_X$ (GeV/c$^2$) 
shown as a function of the mass dimension--$n$ of the associated 
quantum field operator.}

\vspace{0.2cm}

\begin{tabular}{lccccc}
Decay Mode & & &$n$& & \\
           &5&6&7  &8&9 \\
\hline
$\mu^-\gamma$       & $1 \times 10^{18}$ & $1 \times 10^9$ & $1 \times 10^6$ 
                    & $3 \times 10^4$ & $4 \times 10^3$ \\
$\mu^-\pi^0$        & $1 \times 10^{18}$ & $1 \times 10^9$ & $1 \times 10^6$ 
                    & $3 \times 10^4$ & $4 \times 10^3$ \\
$\mu^-\eta$         & $6 \times 10^{17}$ & $7 \times 10^8$ & $8 \times 10^5$ 
                    & $3 \times 10^4$ & $3 \times 10^3$ \\
$\mu^-\gamma\gamma$ & $9 \times 10^{17}$ & $9 \times 10^8$ & $9 \times 10^5$ 
                    & $3 \times 10^4$ & $4 \times 10^3$ \\
$\mu^-K^0_L$        & $5 \times 10^{17}$ & $7 \times 10^8$ & $8 \times 10^5$ 
                    & $3 \times 10^4$ & $3 \times 10^3$ \\
$\mu^-K^0_S$        & $4 \times 10^{17}$ & $6 \times 10^8$ & $7 \times 10^5$ 
                    & $2 \times 10^4$ & $3 \times 10^3$ \\
$e^-\gamma$         & $5 \times 10^{18}$ & $2 \times 10^9$ & $2 \times 10^6$ 
                    & $5 \times 10^4$ & $5 \times 10^3$ \\
$e^-\pi^0$          & $4 \times 10^{18}$ & $2 \times 10^9$ & $2 \times 10^6$ 
                    & $4 \times 10^4$ & $5 \times 10^3$ \\
$e^-\eta$           & $9 \times 10^{17}$ & $9 \times 10^8$ & $9 \times 10^5$ 
                    & $3 \times 10^4$ & $4 \times 10^3$ \\
$e^-\gamma\gamma$   & $9 \times 10^{17}$ & $9 \times 10^8$ & $9 \times 10^5$ 
                    & $3 \times 10^4$ & $4 \times 10^3$ \\
$e^-K^0_L$          & $6 \times 10^{17}$ & $7 \times 10^8$ & $8 \times 10^5$ 
                    & $3 \times 10^4$ & $3 \times 10^3$ \\
$e^-K^0_S$          & $2 \times 10^{17}$ & $4 \times 10^8$ & $5 \times 10^5$ 
                    & $2 \times 10^4$ & $3 \times 10^3$ \\
$e^-\rho$           & $9 \times 10^{16}$ & $3 \times 10^8$ & $4 \times 10^5$ 
                    & $2 \times 10^4$ & $2 \times 10^3$ \\
$e^-\omega$         & $9 \times 10^{16}$ & $3 \times 10^8$ & $4 \times 10^5$ 
                    & $2 \times 10^4$ & $2 \times 10^3$ \\
$e^-K^{0^\star}$    & $2 \times 10^{17}$ & $4 \times 10^8$ & $6 \times 10^5$ 
                    & $2 \times 10^4$ & $3 \times 10^3$ \\
\end{tabular}
\end{table}

\clearpage

\end{document}